\documentstyle[12pt,epsfig]{article}
\begin{document}
\font\ninerm = cmr9
\def\footnoterule{\kern-3pt \hrule width
\hsize \kern2.5pt} \pagestyle{empty}
\vskip 0.5 cm

\begin{center}
{\large\bf A perspective on Quantum Gravity Phenomenology\footnote{\uppercase{T}hese
notes provided the basis for the ``summary talk" which \uppercase{I}
gave as chairman of the \uppercase{QG}1 session
(``\uppercase{Q}uantum \uppercase{G}ravity \uppercase{P}henomenology")
at the ``10th \uppercase{M}arcel \uppercase{G}rossmann
\uppercase{M}eeting on \uppercase{G}eneral \uppercase{R}elativity"
(\uppercase{R}io de \uppercase{J}aneiro, \uppercase{J}uly 20-26, 2003).}}

\end{center}
\vskip 0.5 cm

\begin{center}
{\bf Giovanni Amelino-Camelia}\\
{\it $^a$Dipart.~Fisica, Univ.~Roma ``La Sapienza'' and INFN Sez.~Roma1\\
P.le Moro 2, 00185 Roma, Italy}
\end{center}

\vspace{1cm}
\begin{center}
{\bf ABSTRACT}
\end{center}

{\leftskip=0.6in \rightskip=0.6in
I give a brief overview of
some Quantum-Gravity-Phenomenology research lines, focusing on studies
of cosmic rays and gamma-ray bursts
that concern the fate of Lorentz symmetry in quantum spacetime.
I also stress that the most valuable phenomenological analyses
should not mix too many conjectured new features of quantum
spacetime, and from this perspective it appears that
it should be difficult
to obtain reliable guidance on the quantum-gravity problem
from the analysis of synchrotron
radiation from the Crab nebula and from the analysis of
phase coherence of light from extragalactic sources.
Forthcoming observatories
of ultra-high-energy neutrinos should provide several opportunities
for clean tests of some simple hypothesis for the short-distance structure
of spacetime. In particular, these neutrino studies, and some
related cosmic-ray studies, should provide access
to the regime $E>\sqrt{m E_p}$.}

\section{Quantum Gravity Phenomenology}
Quantum-gravity research used to be completely detached from experiments.
The horrifying smallness of the expected quantum-gravity effects,
due to the overall suppression by powers of the ratio of the Planck length
($L_p \sim 10^{-35}m$)
versus the characteristic wavelength of the particles involved in the process,
had led to the conviction that
experiments could never possibly help.
But recently there has been a sharp change in the attitude of
a significant fraction of the quantum-gravity community.
This is reflected also by the tone
of recent quantum-gravity reviews
(see, {\it e.g.}, Refs.~\cite{leeREV,ashtREV,crHISTO,carlipREV})]
as compared to the tone of quantum-gravity reviews published up to
the mid 1990s (see, {\it e.g.}, Ref.~\cite{chrisREV,stachasht}).

The fact that the smallness of an effect does not
necessarily imply that it cannot be studied experimentally
is not actually a new idea, and indeed it finds support in
several examples in physics. Even remaining
in the context of fundamental physics there is the noteworthy
example of studies of
the prediction of proton decay within certain grandunified theories
of particle physics. The predicted proton-decay probability
is really small, suppressed by the fourth
power of the ratio between the mass of the proton and the
grandunification scale $[m_{proton}/E_{gut}]^4 \sim 10^{-64}$,
but in spite of this truly horrifying suppression,
with a simple idea we have managed to acquire an excellent sensitivity
to the new effect. The proton lifetime predicted by grandunified
theories is of order $10^{39}s$ and ``quite a few" generations
of physicists should invest their entire lifetimes staring at a single
proton before its decay, but by managing to keep under observation
a large number of protons our sensitivity to proton decay is dramatically increased.

We should therefore focus our attention\cite{polonpap}
on experiments which have something to do with spacetime
and such that there is an ordinary-physics dimensionless quantity
large enough that it could amplify
the extremely small effects we are hoping to discover.
Over these past few years several new ideas for tests of
Planck-scale effects have appeared at an increasingly fast pace,
with a growing number of research groups joining the
quantum-gravity-phenomenology effort.

Among the quantum-gravity-phenomenology research lines the one
which has been so far most extensively developed concerns
the investigation of the fate of Lorentz symmetry in quantum gravity.
The relevant proposals of Lorentz-symmetry tests
focus primarily on
the implications of Planck-scale effects
for the analysis of gamma-ray bursts\cite{grbgac,billetal},
the possible role of the Planck scale in the determination
of the energy-momentum-conservation threshold conditions
for certain particle-physics reaction
processes\cite{kifu,ita,aus,gactp,jaco,seth},
and the possible role of the Planck scale in the evaluation
of particle-decay amplitudes\cite{gacpion,orfeupion}.

Another key area of interest for quantum-gravity phenomenology
is the one of interferometry. Possible signatures of Planck-scale
physics have been considered for matter interferometers\cite{strunz},
for large ``free-mirror" laser-light
interferometers\cite{gacgwi,nggwi},
and for small-size laser-light interferometers whose mirrors
are rigidly connected\cite{laemGWI}.
Moreover, there is a long-term research programme which
focuses on possible Planck-scale-induced departures from
CPT symmetry\cite{huetpesk,kostcpt,emln,floreacpt,gacbucc}.
And together with these most developed quantum-gravity-phenomenology
research lines several other proposals are being considered by
small networks of research groups.

Rather than attempting a comprehensive review, in Section~2
I will use the example of certain tests of Lorentz symmetry
to illustrate the general structure of a quantum-gravity-phenomenology
research line.
The discussion of gamma-ray bursts and ultra-high-energy
cosmic rays that I present in Section~2 focuses on finding
a direct link between one or two simple hypotheses about new properties
of quantum spacetime and certain characteristic new effects.
I will argue that this conservative strategy, in which the analysis
does not mix too many simultaneous assumptions about
the structure of spacetime at the Planck scale,
can provide valuable insight on the quantum-gravity problem.
In Section~3 I consider certain types of observations which
represent tempting opportunities to speculate about possible
implications of quantum properties of spacetime, but require us
to combine several assumptions about the structure of spacetime
at the Planck scale, and I argue that in these cases
it might be
hard to obtain reliable guidance on the quantum-gravity problem.
In the closing section (Section 4) I comment on forthcoming
UHE (ultra-high-energy) neutrino observatories, as
one of our best chances,
for the near future, of enriching significantly the type of
data used in Quantum Gravity Phenomenology. And I will stress that
the relevant UHE neutrinos, just like the UHE cosmic rays considered
in Section 2, can give us access to a ``Planck-scale ultrarelativistic
regime", in which the ratio $E/m$ (energy versus mass of the particle)
is larger than the ratio $E_p /E$, where $E_p$ is the
Planck energy scale ($E_p \equiv 1/L_p \sim 10^{28}eV$).

\section{The fate of Lorentz symmetry in quantum spacetime}
Models based on an approximate Lorentz symmetry, with
Planck-scale-dependent departures from exact Lorentz symmetry,
have been recently considered in most quantum-gravity research
lines, including models based on ``spacetime foam"
pictures\cite{grbgac,garayPRL}, ``loop quantum gravity"
models\cite{gampul}, certain ``string theory"
scenarios\cite{susskind,bertoNC}, and ``noncommutative
geometry"\cite{susskind,gacdsr,dsrnext}.

The most studied characterization
of these Planck-scale departures from Lorentz symmetry
assumes that the energy/momentum dispersion
relations for fundamental particles\footnote{For composites
of several fundamental particles the dispersion relation
could take a very different form. In particular,
if the momentum of the composite is obtained by a simple sum
of the momenta of the composing particles, then the energy-momentum
of a composite formed
by $N$ identical fundamental particles all carrying roughly the
same energy-momentum is governed by the dispersion
relation $E_{TOT}^2 \simeq N^2 E^2 = N^2 \vec{p}^2 + N^2 m^2
+ \eta N^n E^n N^2 \vec{p}^2/(N^n E_p^n)
\simeq \vec{p}_{TOT}^2 + m_{TOT}^2
+ \eta E_{TOT}^n \vec{p}_{TOT}^2/(N^n E_p^n)$.
}
should be modified
\begin{equation}
0 = f(E,\vec{p}^2,m;E_p)
\simeq E^2 - \vec{p}^2 - m^2 - \eta \frac{E^n}{E_p^n} \vec{p}^2
~,
\label{displead}
\end{equation}
where the power $n$ parametrizes one of the possible differences
between alternative quantum pictures of spacetime
(with the cases $n=1$ and $n=2$ usually favoured in the literature),
and essentially $\eta$  parametrizes the precise value
of the scale of departures from Lorentz symmetry, which may of course
be somewhat different from the Planck scale (but on the other hand
one does expect, in order to work within the quantum-gravity premises of
these analyses, that roughly $\eta \sim 1$).

The fact that the literature has focused primarily on this
parametrization of the dispersion relation is
mostly due to its simplicity, which makes it a natural first step in a
phenomenology of Planck-scale departures from Lorentz symmetry.
However, this scenario is also more or less directly connected
with various quantum-gravity proposals.
In Loop Quantum Gravity preliminary results\cite{gampul,leeDispRel}
provide support for the dispersion relation (\ref{displead}).
In some rather popular noncommutative spacetimes\cite{lukieAnnPhy}
one also finds evidence in favour of (\ref{displead}).
In String Theory it appears that the modification of the dispersion
relation is not automatic but emerges in presence of
certain natural background fields\cite{susskind,bertoNC}, and for some
background configurations a dispersion relation of the
type (\ref{displead}) is encountered\cite{bertoNC}.
The role that phenomenological studies
of (\ref{displead})
could have in the overall development of
quantum-gravity research has been stressed in
the most recent reviews
by experts of the field (see, {\it e.g.},
Refs.~\cite{leeREV,crHISTO,carlipREV}).

\subsection{Gamma-ray bursts}\label{grbs}
In principle one could test  (\ref{displead}) by making simultaneous
measurements of energy and (space-)momentum.
This turns out to be rather unpractical, at least
when one is aiming for the needed Planck-scale sensitivity.
It is therefore generally assumed that  (\ref{displead}) should
be tested in combination with some other key kinematic property.
Perhaps the most natural proposal is to study (\ref{displead})
with the additional assumption that the velocity $v$ of
the particle should be still obtainable from the dispersion
relation using $v = dE/dp$, as it happens to be the case
in classical spacetime
both in nonrelativistic (Galilei) physics and in relativistic
(Einstein) physics. By assuming the validity of the relation  $v = dE/dp$
one is essentially only assuming\footnote{Indeed most of the relevant
phenomenological analyses assume the validity of $v = dE/dp$.
But alternatives are being, legitimately, considered by some
authors (see, {\it e.g.},
Refs.~\cite{Kosinski:2002gu,Mignemi:2003ab,Daszkiewicz:2003yr,jurekREV}).
While these studies of alternatives to $v = dE/dp$ rely of a large variety of
arguments (some more justifiable some less) in my own perception a
key issue here is whether quantum gravity leads to a modified
Heisenber uncertainty principle, $ [x,p] =1 + F(p)$, in which
case the relation $v = dx/dt \sim [x,H(p)]$ would not lead
to $v = dE/dp$.}
that it should be possible to
introduce some form of Hamiltonian description of particle
systems with standard Heisenberg commutator
($ [x,p] =1$, $v = dx/dt \sim [x,H]$, where $H$ is the energy/Hamiltonian).
Combining (\ref{displead}) with  $v = dE/dp$ one is led to
a velocity law which at ``intermediate energies" ($m < E \ll E_p$)
takes the form
\begin{equation}
v \simeq 1 - \frac{m^2}{2 E^2} +  \eta \frac{n+1}{2} \frac{E^n}{E_p^n}
~.
\label{velLIVbis}
\end{equation}
Such a modified velocity law can be sensitively studied experimentally
focusing on the fact that,
whereas in ordinary special relativity two photons ($m=0$)
emitted simultaneously would always reach simultaneously a far-away detector,
according to (\ref{velLIVbis}) two simultaneously-emitted
photons should reach the detector at different times
if they carry different energy.

This type of effect
can be significant\cite{grbgac,billetal}
in the analysis of short-duration gamma-ray bursts that reach
us from cosmological distances.
For a gamma-ray burst it is not uncommon to find a time travelled
before reaching our Earth detectors of order $T \sim 10^{17} s$.
Microbursts within a burst can have very short duration,
as short as $10^{-3} s$ (or even $10^{-4} s$), and this means that the photons
that compose such a microburst are all emitted at the same time,
up to an uncertainty of $10^{-3} s$.
Some of the photons in these bursts
have energies that extend at least up to the $GeV$ range.
For two photons with energy difference of order $\Delta E \sim 1 GeV$
a speed difference $\eta \Delta E/E_p$  over a time of travel
of $10^{17} s$
would lead to a difference in times of arrival of
order $\Delta t \sim \eta T \Delta E/E_p \sim 10^{-2} s$, which
is significant (the time-of-arrival differences would be larger than
the time-of-emission differences within a single microburst).

Such a Planck-scale-induced time-of-arrival difference
could be revealed\cite{grbgac,billetal}
upon comparison of the structure of the gamma-ray-burst signal
in different energy channels.
Considering the achievable sensitivities\cite{glast}
one concludes that
the next generation of gamma-ray telescopes,
such as GLAST\cite{glast},
can test very significantly (\ref{velLIVbis})
in the case $n=1$ (whereas the effects found in the
case $n=2$ are too small for GLAST).

An even higher sensitivity to possible Planck-scale
modifications of the velocity law could be achieved
by exploiting the fact that, according to
current models\cite{grbNEUTRINOnew},
gamma-ray bursters should also emit a substantial amount of
high-energy neutrinos.
Some neutrino observatories should soon observe neutrinos with energies
between $10^{14}$ and $10^{19}$ $eV$, and one could, for example, compare
the times of arrival of these neutrinos emitted by
gamma-ray bursters to the corresponding times of arrival of
low-energy photons.
Assuming that some technical and conceptual challenges can be
overcome\footnote{For example, this type of analysis requires
an understanding
of gamma-ray bursters good enough to establish whether there are typical
at-the-source time delays. The analysis would loose much of its potential
if one cannot exclude some systematic tendency of
gamma-ray bursters to emit high-energy
neutrinos with, say, a certain delay with
respect to microbursts of photons (although by combining several observations
from gamma-ray bursters at different distances one could partly
compensate for this possible systematic effect).}
one could use this strategy to test very reliably the case
of (\ref{velLIVbis}) with $n=1$, and even perhaps gain some access to
the investigation of the case $n=2$.

\subsection{UHE cosmic rays}\label{uhecr}
In alternative to the proposals considered in the previous subsection,
in which one investigates the dispersion relation (\ref{displead})
in combination with the relation $v = dE/dp$,
there has also been strong interest
in the possibility of testing the implications of
(\ref{displead}) when combined with the assumption of unmodified
laws of energy-momentum conservation.
With a given dispersion relation and a given rule for energy-momentum
conservation one has a complete ``kinematic scheme" for the analysis
of particle production in collisions or decay processes.
In the case in which one combines (\ref{displead})
with unmodified
laws of energy-momentum conservation the analysis of course involves
the added element of complexity due to the fact that one must
necessarily introduce a preferred class of inertial frames\footnote{It
has been recently realized\cite{gacdsr,dsrnext}
that a dispersion relation of the type of (\ref{displead})
can be adopted without necessarily renouncing to the
equivalence of inertial frames, at the cost of a deformation
of boost transformations (just like one can replace the Galileian $m = p^2/2E$
with the special-relativistic $m = \sqrt{(E^2 - c^2 p^2)/c^4}$ without
renouncing to the equivalence of inertial frames, at the cost of replacing
Galilei boosts with Lorentz boosts).
However, if one insists both on the equivalence of inertial frames
and on a dispersion relation of type (\ref{displead}) the law of
energy-momentum conservation cannot remain unmodified\cite{gacdsr}.}.
The parameters ({\it e.g.} the parameter $\eta$) will take different
values in different inertial frames and therefore in order to
combine meaningfully the limits obtained working in different frames
it is necessary
to transform all the results into limits
applicable in a given inertial frame.
It is customary to adopt as this ``preferred" frame
the natural frame for the description
of the CMBR (Cosmic Microwave Background Radiation).

It has been observed that the combination of
(\ref{displead}) with unmodified
energy-momentum conservation
can significantly affect the threshold requirements for
certain particle-producing processes.
Let us for example consider collisions between
a soft photon of energy $\epsilon$
and a high-energy photon of energy $E$ that creates an electron-positron
pair: $\gamma \gamma \rightarrow e^+ e^-$.
For given soft-photon energy $\epsilon$,
the process is allowed only if $E$ is greater than a certain
threshold energy $E_{th}$ which depends on $\epsilon$ and $m_e^2$.
For $n=1$, combining (\ref{displead}) with unmodified
energy-momentum conservation,
this threshold energy $E_{th}$
is found to satisfy
\begin{equation}
E_{th} \epsilon + \eta \frac{E_{th}^3}{8 E_p}= m_e^2
\label{thrTRE}
\end{equation}
(assuming $\epsilon \ll m_e \ll E_{th} \ll E_p$).
The special-relativistic result $E_{th} = m_e^2 /\epsilon$
corresponds of course to the $\eta \rightarrow 0$ limit
of (\ref{thrTRE}).
For $|\eta | \sim 1$ the Planck-scale correction can be
safely neglected as long as $\epsilon > (m_e^4/E_p)^{1/3}$.
But eventually, for sufficiently small values of $\epsilon$ and
correspondingly large values of $E_{th}$, the
Planck-scale correction cannot be ignored.
For $\epsilon \sim 0.01 eV$
the modification of the threshold is already significant,
and this is relevant for the observation
of multi-$TeV$ photons from certain Blazars\cite{aus,gactp}.

And the process $\gamma \gamma \rightarrow e^+ e^-$
is not the only case in which this type of Planck-scale modification
can be important. There has been strong
interest\cite{kifu,aus,gactp,jaco,gacpion,orfeupion,nguhecr}
in ``photopion production", $p \gamma \rightarrow p \pi$,
where again the combination of
(\ref{displead}) with unmodified
energy-momentum conservation
leads to a modification of the minimum proton energy required
by the process (for fixed photon energy).
In the case in which the photon energy is the one typical of CMBR photons
one finds that the threshold proton energy can be significantly shifted
upward (for negative $\eta$), and this
in turn should affect at an observably large level the
expected ``GZK cutoff" for the observed cosmic-ray spectrum.
Observations reported by the AGASA\cite{agasa} cosmic-ray
observatory provide some encouragement for the idea of
such an upward shift of the GZK cutoff, but the issue
must be further explored.
Forthcoming cosmic-ray observatories, such as Auger\cite{auger},
should be able\cite{kifu,gactp} to fully investigate this possibility.

\section{Limitations of ``cocktail analyses" in the search of \\
quantum-gravity signatures in astrophysics}
I stressed that it would be ideal to test directly
(\ref{displead}), without the need of relying on any
other assumption on properties of Planck-scale physics.
This Quantum-Gravity Phenomenology is trying to provide some
hints for the solution of the quantum-gravity problem
and a test of (\ref{displead}) could be useful from this perspective.
But if our phenomenology mixes (\ref{displead}) with other assumptions
we will only test a certain ``cocktail" of assumptions
for quantum gravity, with an obvious decrease in the
quality of the insight gained.
As mentioned we are unable to test sensitively (\ref{displead}) on its
own, but still we should give priority to tests which
require the fewest and the simplest (most natural) assumptions
in combination with (\ref{displead}).
The assumption of $v=dE/dp$, considered in Subsection~2.1,
and the assumption of unmodified energy-momentum conservation,
considered in Subsection~2.2, are good examples of what could
be a single extra assumption to combine with (\ref{displead}).
Unfortunately in certain observational contexts which at first sight
appear to provide a good chance for Planck-scale sensitivity one then
finds out that a comprehensive phenomenological analysis requires
a combination of several
assumptions about the Planck-scale regime.
Two examples which I consider in this section will
illustrate the limitations that can emerge from relying on such
cocktails of assumptions.

\subsection{Synchrotron radiation from the Crab nebula}\label{sync}
A recent series of
papers\cite{jacoNATv1,newlimit,jaconature,tedreply,carrosync,nycksync,tedsteck}
has focused on the possibility to set limits on Planck-scale modified
dispersion relations focusing on their implications for synchrotron radiation.
By comparing the content of the first estimates\footnote{Ref.~\cite{jacoNATv1}
is at this point obsolete, since the relevant manuscript
has been revised for the published version\cite{jaconature}
and the recent Ref.~\cite{tedsteck} provides an even more
detailed and careful analysis. It is nevertheless useful to consider
this series of manuscripts \cite{jacoNATv1,jaconature,tedsteck}
as an illustration of the inevitable increasing level of complexity
of the analysis that emerges as more and more interplays within the
system of assumptions are taken into account.} produced in
this research line\cite{jacoNATv1}
with the understanding that emerged from follow-up
studies\cite{newlimit,jaconature,tedreply,carrosync,nycksync,tedsteck}
one can gain valuable insight on the risks involved in analyses based on
cocktails of several assumptions about Planck-scale physics.
In Ref.\cite{jacoNATv1} the starting point is the observation
that in the conventional (Lorentz-invariant) description of synchrotron
radiation one can estimate the characteristic energy $E_c$ of
the radiation through a heuristic analysis\cite{jackson}
leading to the formula
\begin{equation}
E_c \simeq {1 \over
R {\cdot} \delta {\cdot} [v_\gamma - v_e]}
~,
\label{omegacjack}
\end{equation}
where $v_e$ is the speed of the electron,
$v_\gamma$ is the speed
of the photon, $\delta$ is an angle obtained from the opening angle between
the direction of the electron and the direction of the
emitted photon, and $R$ is the radius of curvature of
the trajectory of the electron.
Ref.~\cite{jacoNATv1} implicitly relies on several
assumptions\cite{newlimit,tedreply,carrosync,nycksync,tedsteck},
including:
(i) the assumption that both the dispersion relation (\ref{displead})
and the relation $v=dE/dp$ are verified;
(ii) the assumption that the same heuristic derivation
of the synchrotron-radiation
cutoff energy applies exactly also at the Planck-scale,
which in particular requires\cite{newlimit,jaconature,carrosync}
that an ordinary effective
low-energy field-theory description is possible;
(iii) the assumption that the relation between
the ``opening angle" $\delta$
and the energy $E$ of the electron
emitting the radiation is unaffected by the Planck-scale
departures from Lorentz symmetry.

As an opportunity to test the corresponding modification of the
value of the synchrotron-radiation cutoff one can hope to
use some relevant data\cite{jacoNATv1,jaconature} on photons detected from
the Crab nebula.
The observational information on synchrotron radiation being emitted
by the Crab nebula is rather indirect: some of the photons we observe
from the Crab nebula are attributed to sychrotron processes on the basis
of a promising (but unconfirmed) conjecture, and the value of the
relevant magnetic fields is also conjectured (not directly measured).
But let me set aside these (however important) facts about the observational
situation, since I here want to focus on the problems that arise when relying
on ``cocktails of assumptions" (independently of the reliability of the data
which are being considered).
The assumptions on which Ref.~\cite{jacoNATv1}
relies clearly limit the insight
gained through the phenomenological analysis.

Assuming that indeed the observational situation has been properly
interpreted and relying on the additional assumptions (i), (ii) and (iii)
one could basically rule out\cite{jacoNATv1} the case $n=1$
for the modified dispersion relation (\ref{displead}).
However, it was then realized\cite{rob} that the assumptions (i) and (ii)
are not fully compatible: if one sets up dynamics according to the
rules of effective low-energy field theory one cannot assume
the dispersion relation (\ref{displead}) to apply to photons.
At linear order in the Planck length ($n=1$) one can write terms
that modify the dispersion relation for photons but the effect
is then automatically such that it involves a strong helicity dependence:
if right-circular polarized photons satisfy the
dispersion relation $E^2 \simeq p^2 + \zeta p^3$ then necessarily
left-circular polarized photons satisfy the ``opposite sign"
dispersion relation $E^2 \simeq p^2 - \zeta p^3$.
For spin-$1/2$ particles Ref.~\cite{rob} does not appear to
impose upon us a similar helicity dependence but of course in a context in
which photons experience such a complete correlation of the
effect with helicity it would be awkward to assume that instead for
electrons the effect is completely helicity independent.
One therefore introduces two independent parameters $\eta_+$ and $\eta_-$
to characterize the
modification of the dispersion relation for electrons.
In the more recent quantum-gravity analyses of synchrotron radiation from
the Crab nebula\cite{tedsteck}
this realization has led to more prudent claims concerning
the implications of these observations for the idea of modifications
of the dispersion relations with terms linear in the Planck length ($n=1$):
the analysis (as presently formulated) is only relevant for quantum-gravity
scenarios that are compatible with the type of needed low-energy effective
field theory
that is used in the analysis and in those contexts it can only be used
to constrain one of the three parameters ($\zeta$,$\eta_+$,$\eta_-$)
that would characterize the
modification of the dispersion relation in
a  low-energy effective-field-theory setup.

I must stress that, while it is of course legitimate to develop
a quantum-gravity-phenomenology test theory that is formulated in
the effective-field-theory language,
by adopting an effective-field-theory formalism one
can anyway only provide rather limited insight for the
overall effort of quantum-gravity research.
In fact, a significant portion of the quantum-gravity community
is justifiably skeptical about the insight gained
from analyses relevant for the quantum-gravity problem
done within low-energy effective field theory.
In particular, the first natural prediction of
low-energy effective field theory
in the gravitational realm is a value of the energy density
which is some 120 orders of magnitude greater\footnote{And the
outlook of low-energy effective field theory
in the gravitational realm does not improve much through the observation
that exact supersymmetry could protect from the emergence of any energy density.
In fact, Nature clearly does not have supersymmetry at least
up to the TeV scale, and this would still lead to a natural prediction
of the cosmological constant which is some 70 orders of magnitude too high.}
than allowed by observations.
Somewhat related to this ``cosmological constant problem"
is the fact that a description of possible Planck-scale departures from
Lorentz symmetry within effective field theory can only be developed
with a rather strongly pragmatic attitude; in fact,
while one can introduce Planck-scale suppressed effects
at tree level, one of course
expects\footnote{Indeed some studies, notably Refs.~\cite{suda1,suda2},
have shown mechanisms such that within an effective-field-theory
approach loop effects would lead to
inadmissibly large
departures from ordinary Lorentz symmetry.}
that loop corrections would naturally lead to inadmissibly large
departures from ordinary Lorentz symmetry.

Perhaps most importantly,
if we look at the quantum pictures of spacetime that provide
support for the proposal (\ref{displead}), which usually involve
either noncommutative geometry or Loop Quantum Gravity,
at the present time one does not find any encouragement for this type of
low-energy effective-field-theory description.
The noncommutative spacetimes in which modifications of the dispersion
relation are being most actively considered are characterized
by spacetime-coordinate noncommutativity of the
type $[x_\mu,x_\nu ] = i \theta_{\mu \nu}
+ i \rho^\beta_{\mu \nu} x_\beta$, and it is well known that the
construction of quantum field theories in these spacetimes requires
the introduction of several new technical tools, which in turn
lead to the emergence of several new physical
features, even at low energies.
When the matrix $\rho$ is present ($\rho \neq 0$)
we are still struggling in the search of a satisfactory formulation of
a quantum field theory\cite{lukieFT,gacmich}.
For the special case $\rho = 0$ the community has achieved substantial progress
in the development of quantum field theories\cite{dougnekr},
but the results actually show that it is not possible to rely
on an ordinary effective low-energy
quantum-field-theory description. In fact, one finds
a surprising ``IR/UV mixing"\cite{susskind,dougnekr,gianlucaken},
{\it i.e.} the high-energy sector of the theory does not decouple
from the low-energy sector, and this in turn affects
very severely\cite{gianlucaken}
the outlook of analyses based on
an ordinary effective low-energy quantum-field-theory description.
Indeed in the study reported in Ref.~\cite{casto},
which was announced soon after the
first papers on the quantum-gravity implications of synchrotron radiation
from the Crab nebula, a quantum field theory
in noncommutative spacetime with
modified dispersion relation was analyzed focusing on
synchrotron radiation and it was argued that the limitations
suggested by Ref.~\cite{jacoNATv1,jaconature} do not actually apply.

The assumption of availability of an ordinary effective low-energy
quantum-field-theory description finds also no support in Loop Quantum Gravity.
Indeed, so far, in Loop Quantum Gravity
all attempts to find a suitable limit of the theory which can be described
in terms of a quantum-field-theory in background spacetime
have failed.
And on the basis of some recent studies\cite{jpDECO}
it appears plausible that in most contexts in which one
would naively expect a low-energy field theory description
Loop Quantum Gravity might predict a density-matrix
description.

Also worrisome is the assumption that the relation between
the opening angle $\delta$ and the energy $E$ of the electron
emitting the radiation should be unaffected by the Planck-scale
departures from Lorentz symmetry.
It is in fact well established that assuming (\ref{displead})
in the analysis of particle-physics processes
one naturally finds striking modifications of the formulas that relate
the energy-momentum of the incoming particles with
the opening angles between the directions of motion
of the outgoing particles.
For example, the result here discussed in Subsection~2.2,
the modification of the threshold
energy for $\gamma + \gamma \rightarrow e^+ + e^-$,
can be viewed\cite{newlimit} as an effect due to a significant
modification of an opening-angle formula.

This concern for the opening-angle estimate of
Ref.~\cite{jacoNATv1,jaconature,tedsteck}
becomes even more serious considering the fact that
synchrotron radiation can be described in terms
of Compton scattering with the virtual photons of the magnetic field.
Describing the virtual photon as a particle with momentum ${P_*}$
and energy ${E_*}$ one finds that in the
process $e^- + \gamma_{{virtual}} \rightarrow e^- + \gamma$
the opening angle $\phi$ between the outgoing particles
must satisfy the relation
\begin{equation}
cos(\phi) \simeq {2 p_f  E_{\gamma,out} - 2 E_i {E_*} - 2 p_i {P_*}
- ({E_*}^2-{P_*}^2) + { E_{\gamma,out} \over E_f} m_e^2
+ {2 \eta E_f^2  E_{\gamma,out} \over E_{p}}
\over 2 p_f E_{\gamma,out}}
~,
\label{syncangle}
\end{equation}
where $E_i$ ($p_i$) is the energy (momentum)
of the incoming electron, $E_f$ is the energy of the outgoing electron,
and $E_{\gamma,out}$ is the energy of the (real, on-shell) photon
that is emitted.
For negative $\eta$ the Planck-scale correction term
can induce a significant reduction of the opening angle,
which would affect\footnote{From private communications
I infer that Jacobson, Liberati and
Mattingly are not too concerned about this opening-angle issue
(while they are making a dedicated effort of exploration\cite{tedsteck}
of the consistency requirements
that emerge from the use of effective field theory in their analysis).
They do comment briefly on this opening-angle issue in Ref.~\cite{tedreply},
but I must leave to the interested reader the task of assessing whether
those brief comments do provide sufficient reassurance.}
the analysis of Ref.~\cite{jacoNATv1,jaconature,tedsteck}.

\subsection{Phase coherence of light from extragalactic sources
and time quantization}\label{phaseco}
The analysis of Ref.~\cite{lieu}
can provide a second example of a physical context in which
at first sight there appears to be a good chance for
Planck-scale sensitivity but then it emerges that
a comprehensive phenomenological analysis requires
a combination of too many
assumptions about the Planck-scale regime.
The observations that are relevant for the analysis of Ref.~\cite{lieu}
are the ones that provide evidence of a good phase coherence of
light from extragalactic sources.
And the key objective of the analysis reported in Ref.~\cite{lieu}
is a test of the possibility that time may be fuzzy/quantized at the
Planck scale ($\tau_p$ quantization, with $\tau_p \sim 10^{-44}s$ the Planck
time).

I should mention parenthetically that the analysis of Ref.~\cite{lieu}
has been criticized at a merely computational level, especially
for what concerns the way in which nonsistematic effects
were combined\cite{ngCRITlieu} and the nature of the conventional-physics
processes that lead to
phase coherence of
light from extragalactic sources\cite{critLIEU2}.
Just like I set aside in the previous subsection the concerns for
the nature of the data on synchrotron radiation from the Crab Nebula,
here I want to set aside this type of skepticism toward the analysis
reported in Ref.~\cite{lieu}.
I intend to focus on the fact that in the phenomenological analysis presented in
Ref.~\cite{lieu}
the hypothesis of Planck-scale time quantization is not tested directly,
but rather it is combined with a rich cocktail of
assumptions, including: (j) that time quantization should be accompanied
by a corresponding level of distance quantization, {\it i.e.} a combined
measurement of space-position and time should be affected
by $\delta x > L_p$ and $\delta t > \tau_p$  uncertainties,
(jj) that a framework hosting
these Planck-scale uncertainties in time and space-position
should necessarily also predict irreducible uncertainties
for energy and space-momentum measurement of the
type $\delta E > E^2/E_p$ and $\delta p > p^2 / E_p$,
and (jjj) that one should also necessarily have a modification
of the dispersion relation roughly of the type (\ref{displead}).

While each of these conjectured features for quantum spacetime
are individually plausible, it is rather ``optimistic" to expect that
all of them should be realized in the correct quantum-gravity theory.
And actually some results in the literature show that
there is no absolute link between the
assumptions (j), (jj) and (jjj).
Once again I can mention some results obtained in the study
of spacetime noncommutativity
of the type $[x_\mu,x_\nu ] = i \theta_{\mu \nu}
+ i \rho^\beta_{\mu \nu} x_\beta$, which is one of the most popular and
simplest ways to introduce spacetime quantization.
A wide body of literature (see, {\it e.g.},
Refs.~\cite{susskind,lukieAnnPhy,gacmich,dougnekr,majrue,gacmaj,wess})
shows that in these quantized spacetimes
energy and momentum, when properly introduced,
are not affected by any minimum-uncertainty conditions,
contrary to the assumption (jj).

Moreover, for the most popular choices of the matrices $\theta$ and $\rho$
the emerging spacetime quantization does not lead
to the conclusion that a combined
measurement of space-position and time should be affected
by $\delta x > L_p$ and $\delta t > \tau_p$  uncertainties.
For example the so-called $\kappa$-Minkowski spacetime
is characterized by $[x_j , x_l] = 0~,~~~[x_j , t] = i x_j /\kappa$
(where $\kappa$ is a noncommutativity scale with dimensions of mass)
and is therefore fully compatible with the possibility of assigning
sharp values to the space coordinates $x_j$ (while indeed the time
coordinate $t$ is subject to a meaningful
Planck-scale discretization\cite{gacmaj}).

\section{The Planck-boost regime}
Together with a brief review and reanalysis of existing areas of
interest in quantum-gravity phenomenology, I like to include in these
notes also some remarks on a previously unnoticed opportunity
which I see for this subject.
I can do this while remaining in the context of the research
line that investigates the fate of Lorentz symmetry in quantum spacetime,
on which I mainly focused throughout these notes.

My point originates from the realization that
some rather different ``amplification mechanisms" are at work in
the various processes I considered. For example, the threshold anomalies
can get large when the two colliding particles have a large
energy difference (one is soft and one is very hard), {\it i.e.}
when the laboratory frame is highly boosted with respect to the
center-of-mass frame. The time-of-travel analyses, which I considered
in discussing gamma-ray bursts, get larger at higher particle energy,
and for a massive particle they are therefore more significant if the
laboratory frame is highly boosted with respect to the rest frame of the
particle.

My discussion was mainly focused on a specific model of Planck-scale departures
from Lorentz symmetry, and it is clearly strongly model dependent (if the model is
changed one can expect even sizeable changes in the nature and magnitude of the
effects). But there is one aspect that might have more general, nearly model-independent,
validity: the new Planck-scale effects become significant for high boosts
with respect to the center-of-mass (or particle-rest) frame.
Within a specific model a detailed analysis is needed in order to establish
which type of high boosts are sufficient for a significant size of the effects.
But it would be useful to have a more general intuition for a large-boost
regime that is of interest from a quantum-gravity perspective.
For example, in studies of the short-distance structure of spacetime
different pictures lead to different expectations for the distance scale
where the nonclassical features become relevant, but there is
a distance scale which is perceived to be intrinsically of interest
from a quantum-gravity perspective: when the processes involve distance scales
of the order of the Planck length most researchers share the expectation
that new effects should be present, quite independently of the specific models
that different researchers are pursuing.
For boosts we do not yet have a similar intuition. The Planck scale is interpreted
equivalently as a length or energy scale, but there is no common expectation
of a characteristic size of boosts that corresponds to the Planckian regime.
To the idea of a large boost most researchers associate the image
of particles with velocity ``close to $1$", but no standard measure
of ``how close is close enough" has been adopted.
Distance scales are small enough
to be potentially sensitive
to Planck-scale effects when they are of the order of the Planck length.
Is there a model-independent way to describe
a particle's velocity as ``high enough to be potentially sensitive
to Planck-scale effects"?

It seems to me that the Planck length also allows to introduce an
objective reference for the magnitude of boosts, at least in
certain contexts. Take in particular a particle of rest energy $m$.
By going from the rest frame to some boosted frames the same particle
will carry a frame-dependent energy $E$, and the size of the energy
of the particle measures the magnitude of the boost needed
to connect the laboratory frame to the rest frame.
In ordinary special relativity $E/m = \cosh (\xi)$,
where $\xi$ is the rapidity of the laboratory frame with respect to
the rest frame.
So energy can measure, in an appropriate sense, the magnitude of some relevant
boosts.
And I observe that the availability of the Planck scale allows to
introduce two different regimes: the low-boost regime $E/m < E_p/E$
and the high-boost regime $E/m > E_p/E$.
In principle one can consider even a series of energy/boost values
that get us deeper and deeper in the Planck regime:
$m  < E < \sqrt{m E_p}$,
$ \sqrt{m E_p}  < E < (m E_p^2)^{1/3}$,
$(m E_p^2)^{1/3}  < E < (m E_p^3)^{1/4}$, $\dots$.

The key point of this simple observation is that it suggests
that, while normally one refers to the Planck regime as the regime of
energies close to the Planck energy, there is a sense in which access
to the Planck regime could be characterized as a requirement for
a combination of particle energy, particle mass and Planck scale.

There is also a rather amusing quantitative observation that I can report
on this point. The standard estimate of the mentioned GZK scale for
cosmic-ray observations is of order $5 \cdot 10^{19} eV$, which
happens to
be\cite{dsrgzk} just above the
scale $\sqrt{m_{proton}E_p} \sim 3 \cdot 10^{18} eV$
for a cosmic-ray proton. One could therefore suspect that the anomalies
reported by the AGASA\cite{agasa} observatory for the cosmic-ray spectrum
(if at all to be trusted) might reflect some new physics
in the high-boost/high-velocity regime $E > \sqrt{m E_p}$.
Of course a minimum requirement for justifying interest in
this numerological observation is confirmation by the mentioned
forthcoming Auger data. But it is noteworthy that, in case it is
needed, cosmic rays are not the only opportunity to access
the high-boost regime.
In fact, as mentioned,
some neutrino observatories should soon observe neutrinos with energies
between $10^{14}$ and $10^{19}$ $eV$, and although the precise values
of the neutrino masses remain undetermined, on the basis of the
present upper limits one can expect that these data
will be sufficient to explore the high-boost regime for neutrinos
(for example $ \sqrt{m E_p} < 10^{14} eV$ only requires that $m < 1 eV$).
By combining the information from observations of UHE cosmic rays and
observations of UHE neutrinos we might gain in the near future some
access to the Planck-boost regime.

\section*{Acknowledgments}
I am grateful to A.~Grillo, T.~Jacobson, T.J.~Konopka, S.~Liberati, D.~Mattingly,
D.~Sudarsky, L.~Urrutia,
who were kindly available for discussions while I was preparing
this manuscript.

\end{document}